\documentclass[
twocolumn,
]{ceurart}

\usepackage{algorithm}
\usepackage{algpseudocode}
\usepackage{hyperref}

\usepackage{multirow}
\usepackage{caption}
\usepackage{tikz}
\usepackage{amssymb}

\usepackage{listings}
\begin{document}

\copyrightyear{2024}
\copyrightclause{Copyright for this paper by its authors.
  Use permitted under Creative Commons License Attribution 4.0
  International (CC BY 4.0).}

\conference{ }

\title{ Punch Out Model Synthesis: A Stochastic Algorithm for Constraint Based Tiling Generation}

\author[1]{ Zzyv Zzyzek }[%
  orcid=0009-0005-2175-1166,
  url=https://zzyzek.github.io/,
]

\begin{abstract}
  As an artistic aid in tiled level design, Constraint Based Tiling Generation (CBTG) algorithms can help
  to automatically create level realizations from a set of tiles and placement constraints.
  Merrell's \textit{Modify in Blocks Model Synthesis} (MMS) and Gumin's \textit{Wave Function Collapse} (WFC) have been proposed
  as Constraint Based Tiling Generation (CBTG) algorithms that work well for many scenarios but have limitations
  in problem size, problem setup and solution biasing.
  We present Punch Out Model Synthesis (POMS), a Constraint Based Tiling Generation algorithm,
  that can handle large problem sizes, requires minimal assumptions
  for setup and can help mitigate solution biasing.
  POMS attempts to resolve indeterminate grid regions by trying to progressively realize sub-blocks, performing a
  stochastic boundary erosion on previously resolved regions should sub-block resolution fail.
  We highlight the results of running a reference implementation on different tile sets and
  discuss a tile correlation length, implied by the tile constraints, and its role in choosing
  an appropriate block size to aid POMS in successfully finding grid realizations.
\end{abstract}

\maketitle

\newcommand{\specialcell}[2][c]{\begin{tabular}[#1]{@{}l@{}}#2\end{tabular}}
\newcommand{\specialcellCenter}[2][c]{\begin{tabular}[#1]{@{}c@{}}#2\end{tabular}}

\section{Introduction}

\subsection{Overview}

\begin{figure*}[ht]
  \includegraphics[width=\textwidth]{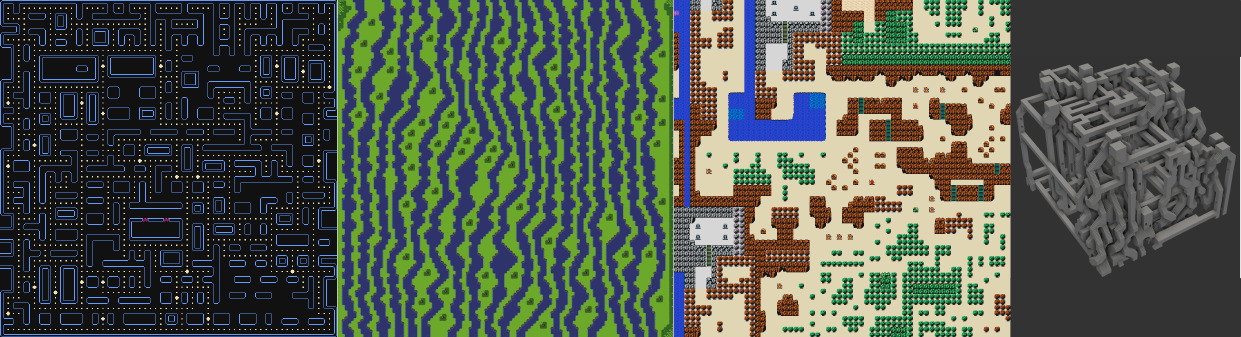}
  \caption{Examples outputs of Punch Out Model Synthesis (POMS) run on different tile sets. From left to right, the \textit{Pill Mortal} tile set, the \textit{Forest Micro} tile set, the \textit{Overhead Action RPG Overworld} tile set and the \textit{Brutal Plum} tile set}
  \label{fig:teaser}
\end{figure*}

We present Punch Out Model Synthesis (POMS), an algorithm that works on a regular 2D or 3D
grid to find a tile placement realization subject to pairwise tile constraints in each grid direction
($\pm X, \pm Y, \pm Z$).

POMS is a grid level stochastic Constraint Based Tiling Generation (CBTG) algorithm whose primary benefits are:

\begin{itemize}
  \item Requires minimal assumptions on initial setup state
  \item Has resources that scale primarily with block size and not grid size
  \item Can reliably find realizations on arbitrarily sized grids with tile constraints that have finite correlation length
\end{itemize}

and primary drawbacks that:

\begin{itemize}
  \item Has limited success for tile constraints that have long range correlation length
\end{itemize}

Further, we:

\begin{itemize}
  \item Provide the results of running a reference implementation on a variety of tile sets (Section 4)
  \item Explore the Tile Arc Consistent Correlation Length (TACCL), a heuristic for tile correlation length,
        that is used to inform the block size choice and solution strategies for a variety of tile sets (Section 4)
\end{itemize}

Here, a grid realization is a single tile assignment per grid cell that respects the
tile constraints.

POMS works by initially setting the grid to an indeterminate state and progressively realizing
sub blocks of the grid.
Block boundary edges that fall within the larger grid body are \textit{pinned} so that, should a
block realization succeed, the block can be re-integrated back into the larger grid.
Should block level realization fail, depending on the type of block realization failure,
either the failed block region is set to an indeterminate state
or the block region is restored to its previous state and all realized region boundaries within the grid
are considered for \textit{erosion} by probabilistically
setting them to an indeterminate state.

POMS is a stochastic algorithm because of the reversion step.
Undoing previous cell realizations, via
block region reversion or erosion, are done in the hopes of
removing a localized contradiction.
Any expectation of progress for POMS primarily comes from choosing an
appropriate block size.
Conceptually, correlation length is the influence that a cell tile choice
has over other cell tile options at distant locations.
In this paper, we attempt to quantify an aspect of correlation length and use it
to inform the block size choice.

If there is a finite length of correlation that one cell's tile choice has with another,
any contradiction that might appear during the course of resolution are localized to a region.
Reverting the region around the contradiction allows for another attempt at finding a
realization without destroying
the bulk of pre-existing realizations elsewhere in the grid.
Under some conditions of configuration randomness, and for many tile constraints,
resolved cells located far enough away from each other have little or no influence over one another.
The correlation distance informs the block size choice as block sizes chosen large enough allow
for the tile values in the middle cells of a block to be chosen independent of any tiles fixed on the boundary.

Expecting center tile choices to be independent of boundary tile values is, in general, unreasonable as
the CBTG problem is known to be NP-Complete.
Even under random configuration assumptions, the issue can be further complicated if the correlation length is not
finite, or finite but large.
Even though the general problem is likely intractable, or the finite correlation length assumption is
violated, many tile constraints are under constrained and the ability of POMS to overcome local constraints
provides enough capability to find realizations.

Unfortunately, for many tile sets, simple global constraints, tile distribution or sparse initial configuration restrictions
are enough to
confound POMS into failing to find a realization.
Different choices for block scheduling policies, 
different block resolution algorithms
and other parameter choices
can help mitigate these shortcomings and will be briefly addressed (Section 4).

\subsection{Definitions}

We discuss some preliminary ideas before describing details
of the Punch Out Model Synthesis (POMS) algorithm.

A \textit{grid} is a collection of \textit{cells} in the shape of a rectangular cuboid,
of size $N_x \cdot N_y \cdot N_z$, ($N_x, N_y, N_z \in \mathbb{N}$).

Each cell in the grid is a variable whose domain is $D$ \textit{tiles} ($D \in \mathbb{N}$).
Values in neighboring cells are subject to a set of provided \textit{tile constraints}.
%
%
Here, the set of tile constraints is pairwise, and only nearest neighbor in each of the major grid dimensions ($\pm X, \pm Y, \pm Z$).

A tile at a grid cell location is said to have \textit{support from a direction} if there is at least
one tile in the neighboring grid cell direction that respects the tile constraints.
A tile, at a grid cell location, \textit{has support} or \textit{is supported} if a tile has support in each direction.

A cell is said to be \textit{resolved} if there is only one tile present and the tile has support from its neighbors.
Should every cell in the grid be resolved, the grid is said to be resolved.
Should a cell hold no tiles, the grid is then said to be in a \textit{contradictory} state.

The set of $D$ tiles is called the \textit{tile domain}.
The set of available tile options at a cell is called the \textit{cell's tile domain} and represents the
available tiles that can be placed at a given cell location.

The problem of Constraint Based Tiling Generation (CBTG) is to
find a single tile assignment to each grid cell location subject to the tile constraints.
That is, resolve every cell in the grid.

It is sometimes desirable to pick out a sub region from a grid for special consideration.
Here we identify a \textit{block} as a sub region of cells from the grid.
Our concern is with blocks that are rectangular cuboid in shape,
of size $M_x, M_y, M_z \in \mathbb{N}$,
and that can be smaller than the grid size ($1 \le M_x \le N_x, 1 \le M_y \le N_y, 1 \le M_z \le N_z$).

If every tile in every cell in a block is supported, the block of cells is said to be in an \textit{Arc Consistent} (AC) state.
That is, subject to the tile constraints, if every tile in every cell in a block region has at least one valid neighboring tile, the block region
is said to be in an arc consistent state.

One method of attempting to put a block region into an arc consistent state is to remove tiles that have no support from the list of permissible
tiles at a cell location.
Each tile removed can have a cascading effect by potentially causing a tile in a neighboring cell to be unsupported.
The repeated process of removing unsupported tiles throughout a block region until a contradiction is encountered or the block region
is in an arc consistent state is called \textit{constraint propagation}.
Constraint propagation
can be used as the basis for a Constraint Based Tiling Generation (CBTG) solver.

In this paper, we define two distinct classes of solvers that we call \textit{block level} solvers and \textit{grid level} solvers.
We define a \textit{block level} solver as an algorithm that
keeps full state of the block it is trying to solve by propagating constraints and maintaining arc consistency
throughout its run.
A \textit{grid level} solver need not keep full state of the grid and often will only keep minimal information about whether
a grid cell is resolved or is in an indeterminate state.
A block level solver typically needs more resources as, for example,
it might maintain a memory intensive data structure associated
with maintaining arc consistency.

POMS is a grid level solver with one of its input parameters designated to specify
which underlying block level solver to use.
In this paper we use \textit{Breakout Model Synthesis} (BMS), a stochastic block level solver
introduced in Hoetzlein's \texttt{just\_math} project  \cite{Hoetzlein_2023}.
For completeness, pseudo-code for \textit{Breakout Model Synthesis} is given in Appendix A.

Note that since  POMS is a grid level solver, the grid can be in an arc \textit{inconsistent}
state during the course of the algorithm.
This poses no problem in and of itself as the block level solver will attempt to put the block
in an arc consistent state while trying to make progress towards a fully resolved
grid.

\section{Related Work}

To our knowledge, Merrell was the first to introduce the modern formulation of Constraint Based Tiling Generation (CBTG)
\footnote{The term \textit{Constraint Based Tile Generators} was coined by Adam Newgas \cite{BorisTheBrave_cbtg_2021} and has been adopted in this paper,
with slightly different wording, as the name for the specialization of the more general Constraint Satisfaction Problem.}
\cite{Merrell_2007, Merrell_2009}.
Merrell introduced a Modify in Blocks Model Synthesis (MMS) algorithm that starts with a fully resolved grid
and applies a one shot block level constraint solver on sub-blocks within the grid.

Merrell noticed that the block level algorithm undergoes a phase transition of solvability
\footnote{ Note that the phase transition is only for \textit{fallible} models. \textit{Infallible} models will be discussed briefly later in this section. }
, with a decreasing probability
for successful block resolution as block size increases.
Instead of attempting to resolve a large grid in one try, the MMS algorithm progressively resolves sub-blocks and re-integrates them back into the grid.
If a block level resolution fails, the block is discarded without altering the grid and another block is chosen.

For many problems, MMS is ideal as it always keeps a full resolution of the grid throughout its run.
Merrell introduced a sequential overlapping schedule for block choice and, for some suitable assumptions
on block size and underlying distribution, the mixing time can be quick, requiring only a few passes through the grid.

Unfortunately, MMS has two major drawbacks, the second of which Merrell noticed and discussed in his thesis:

\begin{itemize}
  \item The requirement of an initially resolved state to start MMS might be difficult to achieve, either in a practical
        or theoretical sense.
  \item Features bigger than the chosen block size will be missed by MMS as there is no way to realize larger features
        through a single block level alteration.
\end{itemize}

Many of the tile sets and tile constraints that Merrell provides in \cite{Merrell_2007, Merrell_2009} have an ``empty'' tile that 
has itself as an admissible neighbor, creating a situation where the initial fully resolved grid can be easily created
by populating each cell with an ``empty'' tile.
All 2D tile sets and tile constraints presented in this paper do not have a valid resolution with a single replicated
tile and require some amount of knowledge about the tile constraints
to create a fully resolved configuration.

In general, for some tile constraints, finding a class of fully resolved initial configurations might be done through engineering effort.
For example, it might be possible to find a small tileable block that is then able to be replicated through to the whole grid
\footnote{ This method was suggested through personal communication with P. Merrell. }.

The inability to handle certain types of unbounded constraints, such as the implicit top and bottom equal river count constraint that appear in
Wo\'zniak's \textit{Forest Micro} tile set (section 4), is the deeper issue with MMS.
Choosing a small block size for MMS could miss novel features whereas too large of a block size either decreases
the likelihood of block resolution or turns MMS into a block level solver.






Gumin introduced the Wave Function Collapse (WFC) project which improved the block level solver used in MMS
and added facilities for automatic tile constraint deduction from exemplar scenes \cite{Gumin_2016}.
WFC uses a principle of maximum entropy heuristic \footnote{Using a \textit{principle of maximum entropy} implies picking a cell to resolve of \textit{minimum} entropy at each step.}
to choose which cells to resolve.
Though extensions are possible, WFC as presented by Gumin is a one-shot block level solver, giving up should a contradiction be encountered.

Since MMS is a grid level solver, other block level solvers can be used, such as WFC, to resolve underlying blocks,
with modifications added to allow for constraints, boundary conditions and other relevant features.
Merrell provides a comparison between MMS and WFC in \cite{Merrell_comparison_2021}.

Breakout Model Synthesis (BMS) was introduced in Hoetzlein's \texttt{just\_math} project \cite{Hoetzlein_2023}.
Hoetzlein's \texttt{just\_math} project also introduced
the Tile Arc Consistent Correlation Length (TACCL) after noticing that knowledge of the tile correlation length could be
used to create algorithms that took advantage of it.
POMS takes the ideas of tile correlation and the TACCL to inform its stochastic backtracking strategies when applied to the larger grid
without needing the resource requirements that BMS, as a block level solver, would require.



\begin{table}[h]
  \centering
  \begin{tabular}[t]{l|cccc}
      & \textit{WFC} & \textit{BMS} & \textit{MMS} & \textit{POMS} \\
    \hline
    \specialcellCenter{Solver Type} & Block & Block & Grid & \textbf{Grid} \\
    \specialcellCenter{Contradiction \\ \ \ Resilience} & No & Yes & Yes & \textbf{Yes} \\
    \specialcellCenter{Block Step \ \ \ \ \\ \ \ \ \ Consistent} & n/a & n/a & Yes & \textit{\textbf{No}} \\
    \specialcellCenter{Indeterminate \\ \ \ Initial State} & Yes & Yes & No & \textbf{Yes} \\
    \specialcellCenter{Ergodic} & Yes & Yes & No & \textbf{Yes} \\
    \hline
  \end{tabular}
  \caption{\textbf{WFC}: Gumin's Wave Function Collapse \\ \textbf{BMS} : Breakout Model Synthesis \\ \textbf{MMS}: Merrell's Modify in Blocks Model Synthesis \\ \textbf{POMS}: Punch Out Model Synthesis (our algorithm)}
  \label{table:CBTGComparison}
\end{table}

Table \ref{table:CBTGComparison} provides a summary of the differences between WFC, BMS, MMS and POMS.
Here, \textit{contradiction resilience} means that the algorithm can recover should a contradiction be encountered,
\textit{block step consistent} means the algorithm is in an arc consistent state after every block resolution,
\textit{indeterminate initial state} means the algorithm doesn't require a fully resolved initial configuration and
\textit{ergodic} means that, in general, all solution states are possibly to reach.
Note that an assertion of being \textit{ergodic} in this context only means solutions
are possible and does not mean \textit{unbiased} as, depending on the tile constraints or configuration,
solution biasing may occur.
The features of Punch Out Model Synthesis (POMS) in Table \ref{table:CBTGComparison} are highlighted for ease of comparison.

In terms of algorithms to ensure arc consistency, Merrell's implementation of MMS \cite{Merrell_mms_2021} and Gumin's implementation of WFC \cite{Gumin_2016} have used AC3 and AC4.
AC3 is easy to understand and can be performant if tile count is low but quickly suffers as tile count increases \cite{Wallace1993WhyAI}.
AC4 is optimal, in general, but requires large amounts of auxiliary space \cite{Mohr_Henderson_1986}.
The reference implementation of POMS uses AC4 exclusively as tile count is often large (1,000 or more).

CBTGs are a specialization of a more general Constraint Satisfaction Problem (CSP).
Karth and Smith offer some history of CSPs, some common concepts and algorithms in \cite{Karth_Smith_2017, Karth_Smith_2022}.
Of note is Karth and Smith's observation that shallow backtracking does little to help resolve conflicts.


Two areas of research activity for CBTGs are attempts to make infinite CBTG algorithms
and attempts at giving more control over created output.
Kleinberg provides an algorithm for infinite WFC by chaining blocks together \cite{Kleinberg_2019}.
While this can produce large scenes, the tile constraints are conditioned so that failure probability is low and Kleinberg admits
that block resolution can fail without any recourse on how to continue.

Of note is Merrell's discussion of \textit{infallible} models \cite{Merrell_2009}, where the tile constraints are
conditioned so as to never be able to encounter a contradiction.
For infallible models, infinite block chaining is always achievable as there is no possibility of a contradiction occurring.
Though infallible models represent a case that is always solvable,
it's unclear how possible or how difficult it is convert tile constraints from exemplar scenes to infallible models.
In particular, all tile sets considered in this paper are fallible models.

Newgas provides \texttt{Tessera}, a software project that implements WFC along with options for a variety of constraints \cite{Newgas_2021}.
Nie et al. provide an extension to WFC to infinite grids but require the tile set to be \textit{complete} or \textit{sub-complete} \cite{Nie_etall_2023}
which may be difficult for tile sets in the wild.
Cooper introduces \texttt{Sturgeon} that incorporates a mid level API to specify more explicit and longer range constraints as an addition
to WFC \cite{Cooper_2022}.
Though Cooper's \texttt{Sturgeon} program and ideas look promising, the sizes involved are relatively small (40x40 and below)
and it's unclear how well the constraints would work on various tile sets, initial conditions or how well the method would scale
for larger level sizes.

Alaka and Bidarra attempt to offer more control over output level design by considering a user interface to group tiles and
weight individual tile probabilities into user specified regions \cite{Alaka_Bidarra_Rafael_2023}.
Langendam and Bidarra attempt to offer more control over output levels through a mixed initiative graphic tool that offers
the ability to interact with the underlying WFC solver in various ways \cite{Langendam_etall_2022}.

Of note is Lucas and Volz's straight forward application of counting tile frequency and measuring the Kullback-Leibler divergence
to attempt to get a better understanding of how biased the resulting generated map is \cite{Lucas_Volz_2019}.
Karth and Smith use a vector-quantized variational auto encoder (VQ-VAE) to create reduced tile domain maps which can then
be input into WFC to produce novel results \cite{Karth_Smith_2017}.


Though automatic tile constraint creation and constraint based tiling generation are distinct ideas, they are often
bundled together as manually creating tile constraints can be labor intensive and the resulting tile constraints
are an input requirement to constraint based tiling generation problems.
Gumin's \textit{Wave Function Collapse} project highlighted an automatic tile constraint creation from exemplar
scenes \cite{Gumin_2016}.
Many resources exist to explain automatic tile constraint creation in detail but of note are Sherrat's summary in
\cite{Sherratt_2019} and an introduction by Newgas in \cite{BorisTheBrave_wfc_2021}.
We briefly go over automatic tile constraint generation in more detail later in this paper (Section 4).

\section{Algorithm}

\subsection{Punch Out Model Synthesis}

Punch Out Model Synthesis (POMS) works to progressively resolve chosen
blocks within a larger grid.
Blocks that fully resolve are then saved back into the larger grid.
Should blocks not be able to fully resolve, portions of the grid
are reverted back to an indeterminate state, depending on the mode of failure.


\begin{figure*}[ht]
  \centering
  \includegraphics[width=\textwidth]{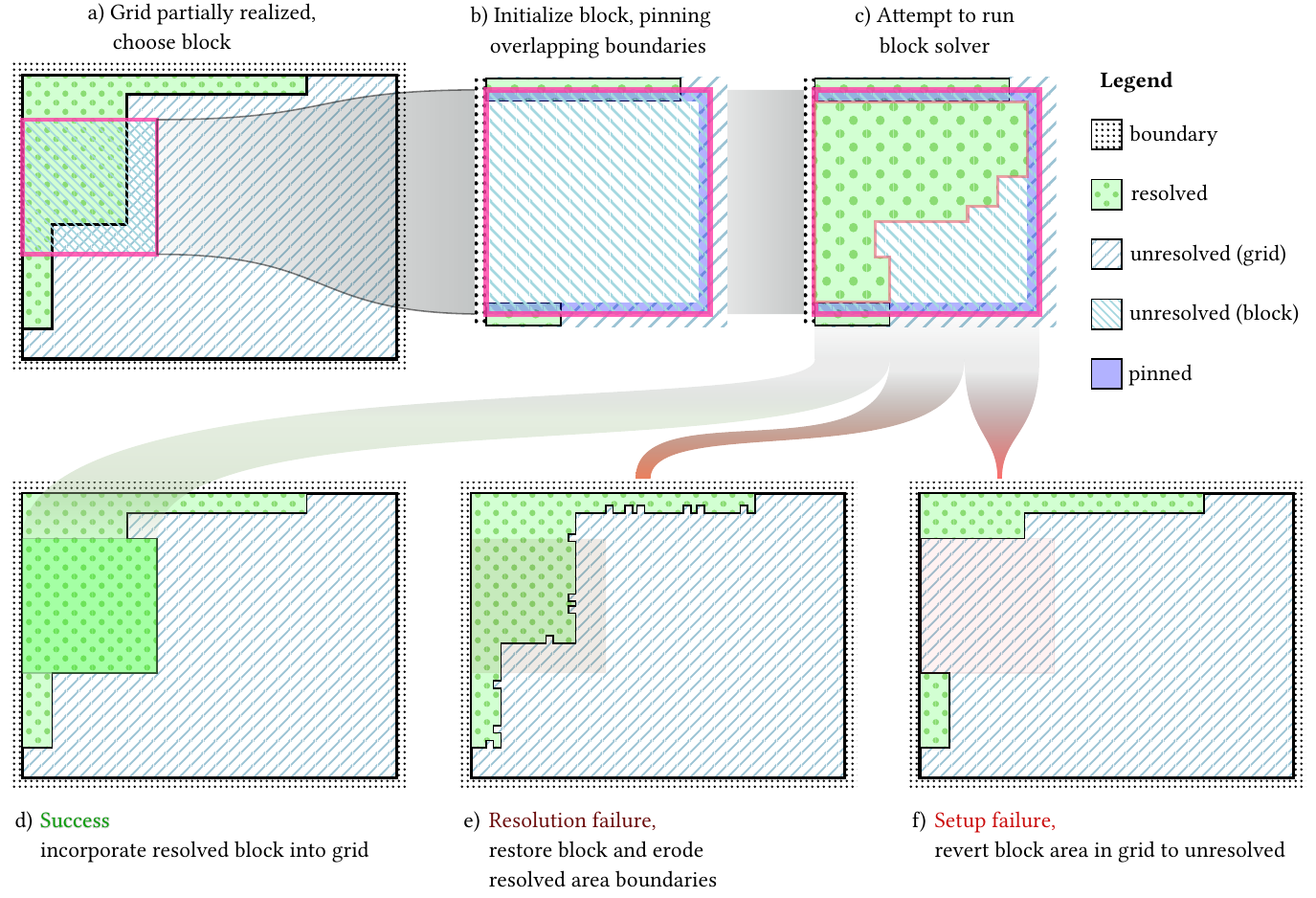}
  \caption{a) A block is chosen in the partially resolved grid, based on a block choice scheduler
  b) Once the block is chosen, the boundary is pinned if not on the a grid boundary and the center put
  into an indeterminate state. c) The block level solver attempts to find a solution for the block,
  with any pinned boundary restrictions d) If successful, the block is incorporated back into the grid.
  e) If the block solver algorithm failed to resolve, after some maximum iteration count, say, then
  the grid is restored to its previous state and resolved boundaries are eroded based on an erosion
  choice scheduler. f) If the block solver algorithm failed to start because the block could not be
  put into an arc consistent state given the tiles pinned on the boundary, the block area in the grid is reverted
  to an indeterminate state.}
  \label{fig:alg}
\end{figure*}


\begin{figure*}[ht]
  \centering
  \includegraphics[width=\textwidth]{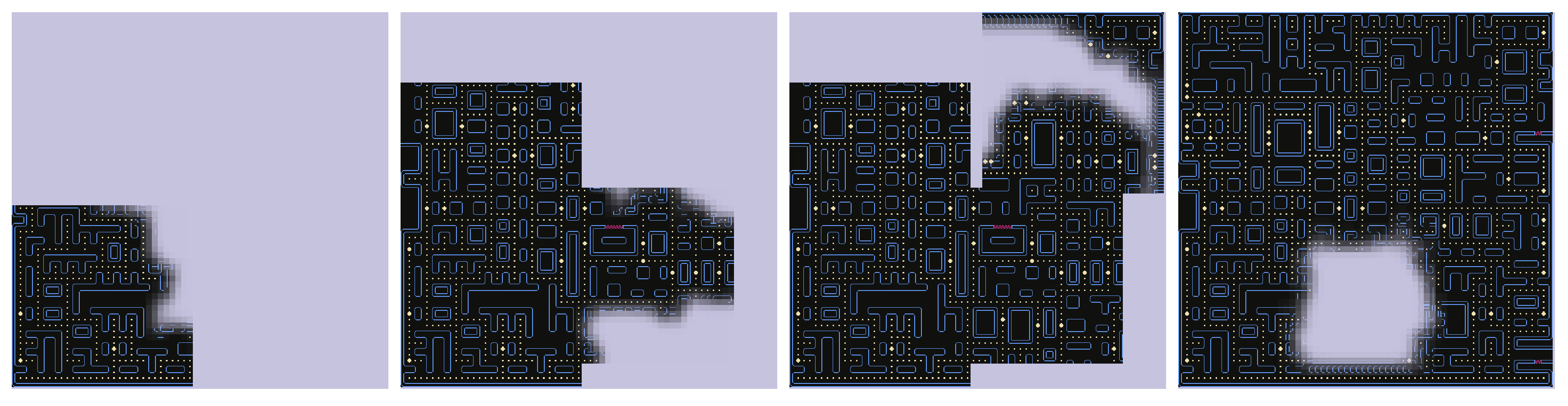}
  \caption{A slideshow of POMS run on the \textit{Pill Mortal} tile set. The block size is 32x32 and the grid size is 64x64 with a block choice policy that chooses block centers uniformly at random from the available unresolved cell locations in the grid. }
  \label{fig:pmrun}
\end{figure*}



\begin{algorithm}
  \caption{Punch Out Model Synthesis}
  \label{alg:POMS}
  \begin{algorithmic}
    \State \textbf{Input,Output:} Grid $G$,
    \State \textbf{Input:} Block Choice Scheduler, $BCS$,
    \State \textbf{Input:} Erosion Choice Scheduler, $ECS$,
    \State \textbf{Input:} Block Solver Algorithm, $A$
    \State Set $G$ to a fully indeterminate state
    \While{ $G$ not fully resolved }
      \State Choose block $B$ in $G$ by querying $BCS$
      \State Initialize $B$ to indeterminate state
      \State From $G$, set and pin $B$ edges not on $G$ boundary
      \State Apply initial setup restrictions to $B$
      \State Run $A(B)$ \Comment{\textit{Attempt to resolve $B$ with $A$}}
      \If{ initial arc consistency is impossible for $B$ }
        \State Set $B$ region to indeterminate state in $G$
      \ElsIf{ $A(B)$ fails to find a full resolution in $B$ }
        \State Erode boundaries in $G$ using $ECS$
      \Else \Comment{\textit{success}}
        \State Copy fully resolved tiles from $B$ into $G$
      \EndIf

    \EndWhile

  \end{algorithmic}
\end{algorithm}


POMS retains a copy of the larger grid, keeping either the fully resolved
tile per cell or an indicator that the cell is in an indeterminate state.
Each round consists of choosing a block from some block choice scheduler 
and doing full block level resolution from some underlying block level algorithm.
In this paper, we only consider Breakout Model Synthesis (BMS) \cite{Hoetzlein_2023}
for the underlying block level resolution algorithm.

Each block chosen has its boundary pinned to the values as they appear in the grid,
except if the block boundary falls on the grid boundary.
Here, \textit{pinning} means setting a cell's tile domain to a certain set of values
and not allowing any constraint propagation to be performed on the cell.
The pinned cell's tile domain can affect a neighboring cell tile but is otherwise not
considered for constraint propagation.
Pinning ensures that the block can integrate into the larger grid by guaranteeing
the boundary conditions of the block match the related regions in the grid.

If the block and grid share a boundary, the tiles at the cell boundaries are not
pinned but are subject to boundary conditions.
Here, we only consider boundary conditions where a fixed tile is used when neighboring
constraints would fall out of grid bounds.

Block regions are chosen by a \textit{Block Choice Scheduler} (BCS) as referenced in Algorithm \ref{alg:POMS}.
From our experimentation, the BCS is problem specific and will be discussed later (Section 4).

Once a block is chosen, the block is initialized by adding the full tile domain of tile values
to every cell in the block, applying setup restrictions and then running an initial constraint propagation
that attempts to put the block in an arc consistent state.
The setup restrictions only allow for tiles to be added, removed
or pinned on an individual cell.
Once setup restrictions are imposed, an attempt is made to run the constraint propagation and
put the block in an arc consistent state.

If the block is successfully put in an initial arc consistent state, the block level resolution algorithm proceeds and attempts to find a fully realized block.
If a fully realized block is found, it's put back into the grid and the algorithm
continues on by choosing another block to process.

If the initial block level arc consistency attempt succeeded but the block level resolution
algorithm failed to find a fully realized block,
boundary tiles of fully resolved regions in the whole grid are probabilistically reverted to an indeterminate state.
The process of probabilistically reverting boundary cell locations to indeterminate states is called \textit{erosion},
with the \textit{probability of erosion} being a tunable parameter that sets the probability that a cell located on a resolved boundary
is reverted.

Should the initial attempt to put the block choice into an arc consistent state fail,
the block region is reverted to an indeterminate state in the grid.
The motivation for reverting the whole block region, as opposed to just relying on erosion,
is that if a block cannot be put into an initial arc consistent state, subject to its pinned
boundary values, there might be a strong contradiction that requires an aggressive
back-off strategy.

Algorithm \ref{alg:POMS} gives pseudo code for the Punch Out Model Synthesis (POMS) algorithm.
Figure \ref{fig:alg} gives an overview of the POMS algorithm, highlighting
a single step.
Figure \ref{fig:pmrun} shows a slideshow of POMS being run on the \textit{Pill Mortal}
tile set for a 64x64 grid size with a 32x32 block size.

The probability of erosion is managed by the \textit{Erosion Choice Scheduler} (ECS) as referenced in Algorithm \ref{alg:POMS}.
We have found that a good choice for ECS is to increase the erosion probability by the number of failed attempts, allowing
for a more aggressive erosion should block resolution become increasingly difficult.
Without the erosion, block level solvers could
perpetually attempt resolution on blocks with identical initial state.
The erosion provides a backtracking mechanism, allowing for a change in initial state
of chosen blocks and removal of implied contradictions that can occur from previously resolved regions in the grid.

Maximum iteration counts can be added so that POMS or the underlying block level resolution
algorithm ($A$) don't loop forever.
The iteration counts and other checks in Algorithm \ref{alg:POMS} have been omitted for brevity.


Since the grid only keeps one value per cell and full constraint propagation is done
on a block level, only a block level's worth of data need be kept in active memory.
Full constraint propagation is done on an individual block level but the block size can be
chosen to be much smaller than the grid, allowing block sizes to be tuned as resources allow.




\subsection{Tile Arc Consistent Correlation Length}

Choosing a block size for POMS is informed by the tile correlation length,
as block sizes below the the tile correlation length run the risk of getting
trapped in a local energy minima without the possibility of escape.
As a heuristic to estimate the tile correlation length,
Hoetzlein's \texttt{just\_math} project \cite{Hoetzlein_2023} proposed
a Tile Arc Consistent Correlation Length (TACCL).

The computation of the TACCL is done as a pre-processing step independent of the POMS algorithm.
The idea is to iterate through each tile in isolation and record the maximum distance constraint propagation reaches when resolving a center cell.

Starting from a small test grid set to an indeterminate state,
the center cell is resolved to a tile and constraint propagation is performed to put the grid
into an arc consistent state.
The size of a bounding box that minimally encompasses every cell whose tile domain was altered during the constraint propagation is then saved.
The grid is then reverted to an indeterminate state and the next tile is chosen to resolve, repeating the process and updating the maximum distances
of the minimum encompassing bounding box for each tile tested.

The TACCL is the maximum extent of the saved bounding boxes from iterating through all tiles.
Algorithm \ref{alg:taccl} provides pseudo-code for the TACCL calculation.

\begin{algorithm}
  \caption{Tile Arc Consistent Correlation Length}
  \label{alg:taccl}
  \begin{algorithmic}
    \State \textbf{Output:} Tile Arc Consistent Correlation Length $L$
    \State \textbf{Input:} Block Size $(M _ x, M _ y, M _ z)$
    \State Create a block $B$ of size $(M _ x, M _ y, M _ z)$
    \State Put block $B$ in a fully indeterminate state
    \State $L=1$
    \For { tile every tile $t \in D$ }
      \State $B' = B$
      \State Apply initial setup restrictions to $B'$
      \State Resolve center cell, $c$, in $B'$ to $t$
      \State Make $B'$ arc consistent
      \State Find minimum bounding box, $bbox$, \\ \ \ \ \ \ \ \ \ encompassing altered cells of $B'$
      \If{ $\max _ { x, y, z } bbox  > L$ }
        \State $L = \max _ { x, y, z } bbox $
      \EndIf
    \EndFor
    \State Return $L$
  \end{algorithmic}
\end{algorithm}

For the sake of brevity, no checks are performed in Algorithm \ref{alg:taccl} to determine if the calculated value
is as large as the input block size ($M _ x, M _ y, M _ z$).
In such a case, either the TACCL is unbounded or the test block size is smaller than the TACCL and the test block size would
need to be increased.

The TACCL is meant to estimate the correlation length of an underlying tile constraint set but can be misleading as some
tile constraints will have a finite TACCL even though correlation lengths can be unbounded.
An unbounded TACCL implies an unbounded correlation length but the converse is not true and a finite TACCL
does not necessarily imply a finite correlation length.
The \textit{Brutal Plum} tile set, that has an unbounded correlation length but finite TACCL,
displays this phenomenon and will be discussed later (Section 4).


\section{Results}

In this section, we highlight five tile sets that represent different aspects of the
benefits and pitfalls of the Punch Out Model Synthesis (POMS) algorithm.

We start with a brief review of the automatic tile constraint creation.
Though automatic tile constraint creation is a distinct topic from Constraint Based
Tiling Generation (CBTG), the two are closely related as any CBTG algorithm requires
a tile constraint set to run.

\begin{figure}[h]
  \includegraphics[width=\linewidth]{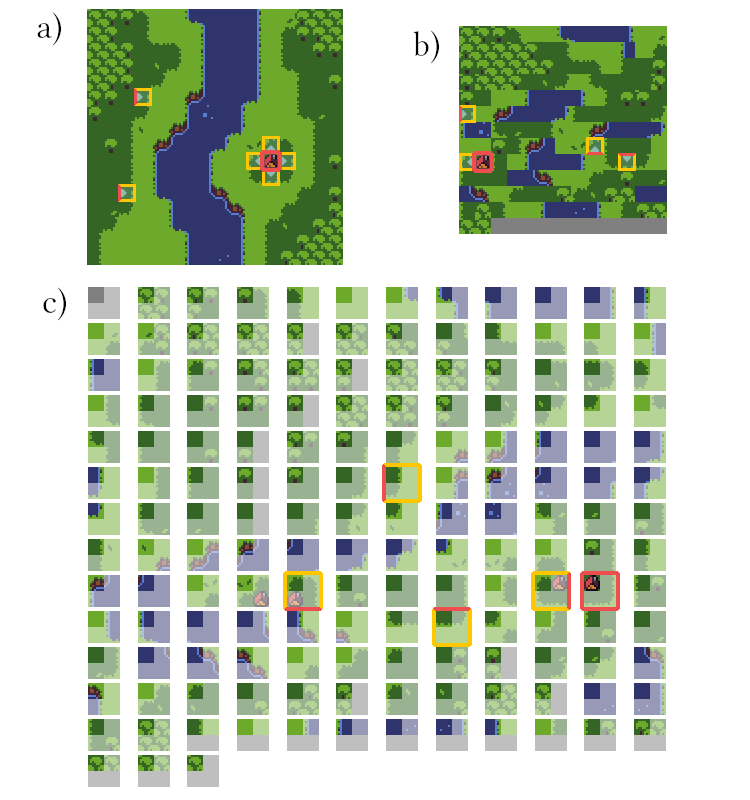}
  \caption{ a) Exemplar image with a single tile and its neighbors highlighted. b) The packed image inferred tile set with the relevant tile highlighted.
            c) The catalog of 2x2 super tiles used to create the tile constraints from the 1x2 tile overlap, suitably rotated. The same tile highlighted in a) and b) has been highlighted here for comparison. }
  \label{fig:rrti_tileset}
\end{figure}


\subsection{Automatic Tile Constraint Creation}

The four 2D tile sets used in this paper had their tile constraints inferred from an exemplar image using
an automated tile constraint generation method.
The exemplar image is split into tiles and a sliding super tile window, of a larger size than the tile itself, is moved over the exemplar image.
The super tiles are deduplicated and checked for overlapping bands to other super tiles.
For every matching, overlapping band, a tile constraint is added, allowing an admissible pairing for the tiles in the
appropriate dimension direction.

Figure \ref{fig:rrti_tileset} shows the exemplar image (\textit{a}), the inferred tile set (\textit{b}) and the complete list of super tiles
for Wo\'zniak's \textit{Forest Micro} tile set (\textit{c}).
A single tile is highlighted in red and its neighbors are highlighted in yellow with red edges corresponding to their neighboring direction.

Figure \ref{fig:rrti_tileset} uses a tile size of 8x8 pixels, with a 2x2 super tile size.
The upper left hand tile is used as the displayed representative tile, which can be seen by comparing the non dimmed tiles in the
super tile list from Figure \ref{fig:rrti_tileset} (\textit{c}) to the packed tile set image (\textit{b}).
A 1x2 overlapping band is tested for equality in each direction, suitably rotated, to find which super tiles neighbor other super tiles.
An interested reader can confirm that there is a 1x2 overlap to the highlighted red super tile to each of its highlighted yellow neighbors
in Figure \ref{fig:rrti_tileset}.

Boundary constraints need special consideration.
One option is to only allow a special ``zero'' tile at grid boundaries, ensuring ``zero'' tile constraints are added
for tile pairs that fall across the edge boundaries of the exemplar image.
This is the option chosen for Wo\'zniak's \textit{Forest Micro} tile set and can be seen in the ``zero'' (gray) tiles present for
the super tiles listed in Figure \ref{fig:rrti_tileset} (\textit{c}).
Another option is to have \textit{wrap around boundary conditions}, with a sliding window that wraps right or up to left or down directions, respectively,
and is the method chosen when creating the tile constraints for LUNARSIGNALS' \textit{Overhead Action RPG Overworld} tile set.

If the tile constraints include the ``zero'' tile at the grid boundaries, this will be denoted as \textit{hard boundary conditions}.


For the automatic tile constraint creation from exemplar images, some artistic input is needed in choosing tile size, window size and boundary conditions.
The tile sizes, window sizes and boundary conditions for the 2D tile sets used in this paper were chosen through inspection and experimentation.
An in depth explanation of automatic tile constraint creation is beyond the scope of this paper and readers are referred to \cite{Gumin_2016, Sherratt_2019, BorisTheBrave_wfc_2021} for further details.

\begin{table}[h]
  \centering
  \begin{tabular}[t]{l|ccccc}
    & \textit{PM} & \textit{OARPGO} & \textit{FM} & \textit{2BMMV} & \textit{BP} \\
    \hline
      Dimension & 2D & 2D & 2D & 2D & 3D \\
      Tile Size (px) & 8px & 16px & 16px & 24px & n/a \\

      \textit{Super Tile} & & & & & \\
      \specialcell{\ \ Window }  & 2x2 & 3x3 & 2x2 & 2x2 & n/a \\
      \specialcell{\ \ Overlap }  & 1x2 & 2x3 & 1x2 & 1x2 & n/a \\
      Tile Count & 190 & 3031 & 159 & 1807 & 90 \\
      \hline

      \specialcell{\textit{Boundary} \\ \ \ \textit{Conditions}} & & & & & \\
      Hard & \checkmark & & \checkmark & \checkmark & \checkmark \\
      Wrap & & \checkmark & & & \\
      \hline

      \specialcell{\textit{Block Choice} \\ \ \ \textit{Scheduler}} & & & & & \\
      Uniform & \checkmark &  & & \checkmark & \checkmark \\
      Diagonal & & & \checkmark & & \\
      Center Out & & \checkmark & & & \\

      \hline
      TACCL (x/y) & 24 & 50/70 & $\infty$ & $\infty$ & 16 \\
      Soften Size & 8 & 12-24 & 8 & 8-24 & 1-8 \\
      Block Size & 32 & 50x70 & 48 & 48 & 22 \\

     \hline
  \end{tabular}
  \caption{\textit{Pill Mortal} (\textbf{PM}) tile set \\ \textit{Overhead Action RPG Overworld} (\textbf{OARPGO}) tile set \cite{LUNARSIGNALS_oarpgo} \\ \textit{Forest Micro} (\textbf{FM}) tile set \cite{ThKaspar_micro} \\ \textit{Two Bit Micro Metroidvania} (\textbf{2BMMV}) tile set \cite{0x72_2bmmv} \\ \textit{Brutal Plum} (\textbf{BP}) tile set}
  \label{table:tilesets}
\end{table}



\begin{figure*}[ht]
  \includegraphics[width=\textwidth]{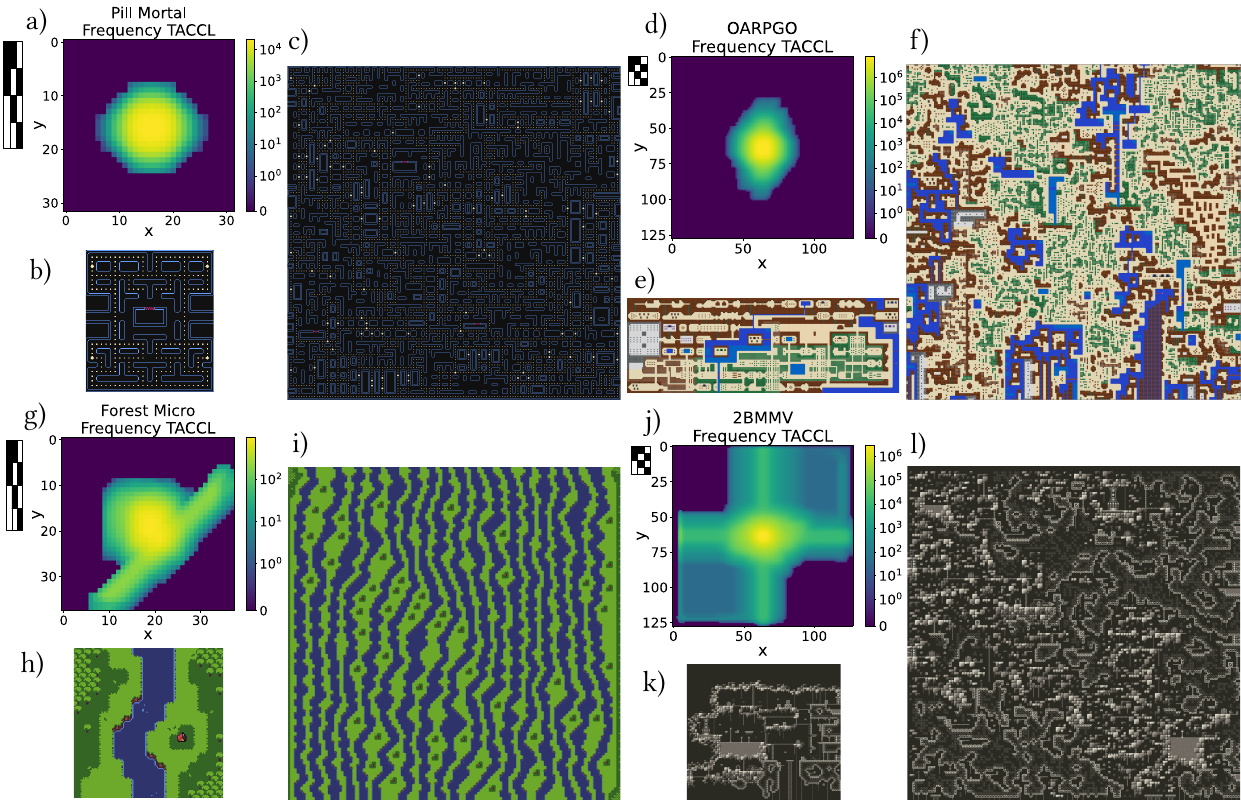}
  \caption{Tile Arc Consistent Correlation Length (TACCL) plots, source exemplar image and example output for four 2D tile sets as listed in Table \ref{table:tilesets}.
           The TACCL, exemplar image and example 64x64 output using a block size of 8x8 for the \textit{Pill Mortal} tile set are shown in a), b) and c) respectively. The TACCL, exemplar image and an example 256x256 output using a block size of 50x70 for LUNARSIGNALS' \textit{Overhead Action RPG Overworld} are shown in d), e) and f) respectively. The TACCL, exemplar image and an example 128x128 output using a block size of 48x48 for Wo\'zniak's \textit{Forest Micro} tile set are shown in g), h) and i) respectively. The TACCL, exemplar image and an example 128x128 output using a block size of 48x48 for 0x72's \textit{Two Bit Micro Metroidvania} tile set are shown in j), k), l) respectively. }
  \label{fig:2dexamples}
\end{figure*}

\begin{figure}[h]
  \centering
  \includegraphics[width=\linewidth]{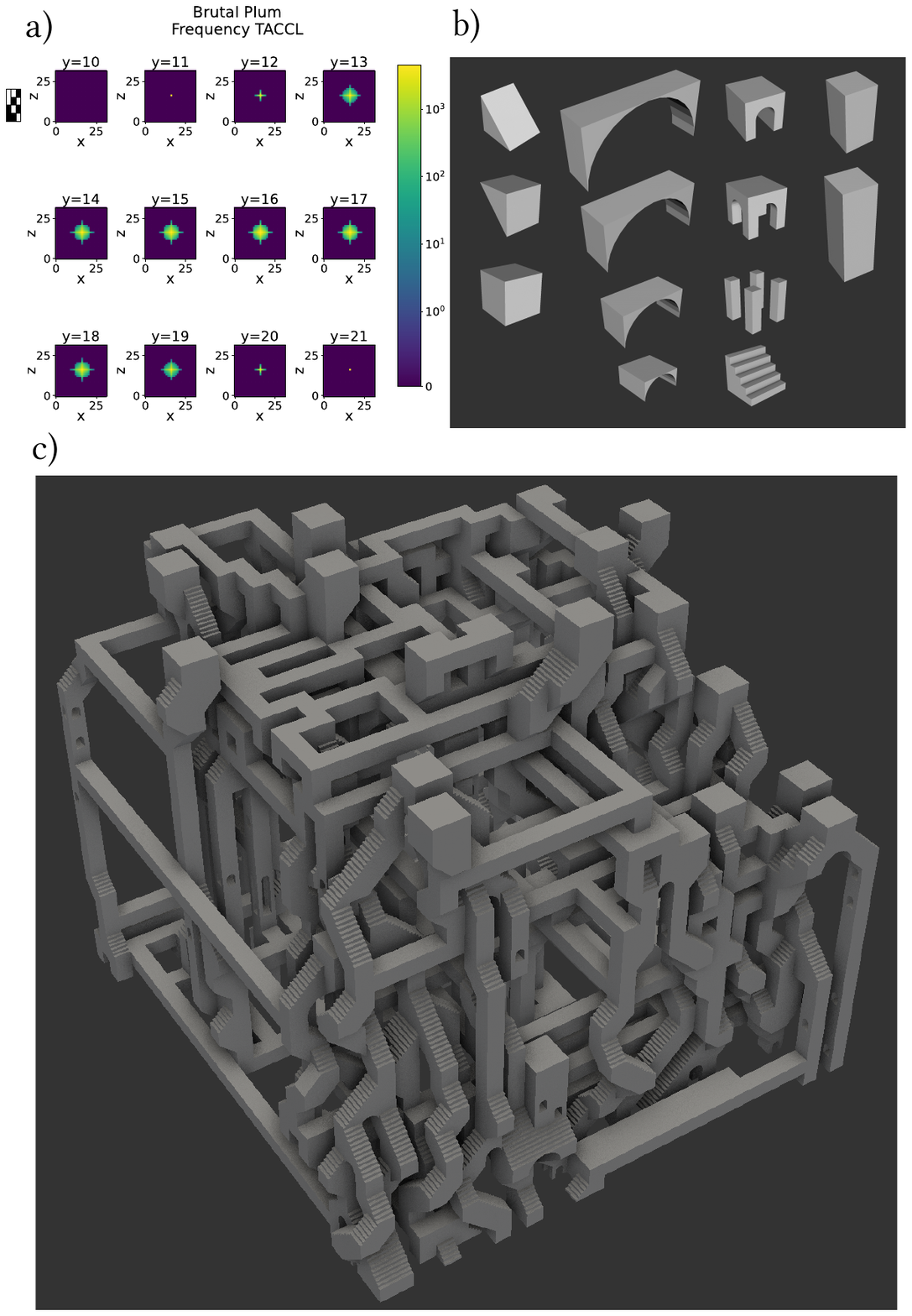}
  \caption{The Tile Arc Consistent Correlation Length (TACCL) plot, source tile set and an example 32x32x32 output for the 3d \textit{Brutal Plum} tile set listed in a), b) and c) respectively. A block size of 22x22x22 was used.}
  \label{fig:brutal_plum}
\end{figure}

Table \ref{table:tilesets} provides summary information for five tile sets, \textit{Pill Mortal}, \textit{Overhead Action RPG Overworld},
\textit{Forest Micro}, \textit{Two Bit Micro Metroidvania} and \textit{Brutal Plum}, highlighted in this paper.
The tile size, super tile window size, super tile window overlap parameter and boundary conditions are provided as they are used in the automatic tile constraint generation
for the resulting tile counts listed.
Also provided are the Block Choice Scheduler (BCS), soften size and blocks sizes that were used to generate the example outputs in Figure \ref{fig:2dexamples} and Figure \ref{fig:brutal_plum}.

Here, the \textit{uniform} BCS denotes a block choice schedule that randomly chooses a block center uniformly from
indeterminate cells in the grid.
A \textit{diagonal} BCS denotes block center choices weighted preferentially to the upper left corner and
\textit{center out} BCS denotes block center choices weighted preferentially towards the center of the grid.

The soften size is specific to the Breakout Model Synthesis (BMS) block level solver and the values shown in Table \ref{table:tilesets} were chosen through experimentation.
We document the soften size for completeness but a more thorough investigation of the soften size's impact on solvability
is beyond the scope of this paper.

We group the 2D tile sets into \textit{bounded} and \textit{unbounded} TACCL, with special consideration for
the \textit{Brutal Plum} tile set.

\subsection{Tile Constraints with Bounded TACCL}

The first two tile sets that appear in Figure \ref{fig:2dexamples}, \textit{Pill Mortal} and LUNARSIGNALS' \textit{Overhead Action RPG Overworld}
(OARPGO) tile set,
have bounded Tile Arc Consistent Correlation lengths (TACCL).

The \textit{Pill Mortal} tile constraints were generated from the exemplar image shown (Figure \ref{fig:2dexamples}, label \textit{b}), using
an 8x8px tile size, hard boundary conditions with a 2x2 super tile window, a 1x2 tile overlap and the upper left hand corner tile as the representative
display tile.
The generated realization in Figure \ref{fig:2dexamples} (label \textit{c}) was created from a POMS run with block size 32x32 on a 128x128 grid.
A uniform block choice scheduler policy was chosen that selected block centers uniformly at random from the set all unresolved grid cells.

LUNARSIGNALS' \textit{OARPGO} tile constraints were generated from the exemplar image shown (Figure \ref{fig:2dexamples}, label \textit{e}), using
an 16x16px tile size, wrap around conditions with a 3x3 super tile window, 2x3 tile overlap and the middle tile as the representative
display tile.
In this case, a 3x3 window was chosen as a smaller 2x2 window was observed to not give good aesthetic results.

When generating outputs using the OARPGO tile set, the outer frame of the grid is pinned to an unresolved state.
This is necessary as the tile constraint set was created with wrap around conditions and has no valid constraints that match the ``zero'' tile
to the rest of the tile domain.

The example 256x256 output (Figure \ref{fig:2dexamples}, label \textit{f})
was created with a block size of 50x70.
A block choice strategy was used that preferentially chose block centers from unresolved grid cells
nearer to the center.
This has the effect of resolving regions from the \textit{center out}, never creating isolated regions that need to be joined and effectively
  ensuring a large contiguous region during the course of the algorithm.
From observation, other block choice strategies were not as successful as choosing block centers with a grid center bias.

It should be noted that without wrap around boundary conditions, the OARPGO tile set would have unbounded TACCL.
Further, without wrap around boundary conditions POMS has trouble finding solutions as the
grid boundary region is non trivial for this tile set.
The success of the center out block choice strategy, the failure of other block choice strategies and the unbounded TACCL of
non wrap around boundary conditions, could suggest that the bounded TACCL does not capture some longer range or unbounded constraints
  embedded within the wrap around OARPGO tile set constraints.

Of note is the ``stuttering'' effect that happens with many features being repeated in a linear direction.
For example, there are long regions of vertical rocks surrounded by water tiles that continue on downward until the
pinned boundary is hit.
This effect becomes even more pronounced when other tile weightings are used.
We suspect this is because of a certain pattern preference that then gets re-enforced
by a surrounding structure.
We make note of this effect but otherwise don't investigate further in this paper.

\subsection{Tile Constraints with Unbounded TACCL}

The last two tile sets in Figure \ref{fig:2dexamples}, Wo\'zniak's \textit{Forest Micro} tile set and 0x72's \textit{Two Bit Micro Metroidvania} tile set,
both have unbounded Tile Arc Consistent Correlation Lengths (TACCL).

The tile constraints for Wo\'zniak's \textit{Forest Micro} tile set were generated from the exemplar image shown (Figure \ref{fig:2dexamples}, label \textit{h}), using
an 16x16px tile size, hard boundary conditions with a 2x2 super tile window, a 1x2 tile overlap and taking the upper left hand corner tile as the representative
display tile.
The generated realization in Figure \ref{fig:2dexamples} (label \textit{i}) was created from a POMS run with block size 48x48 on a 128x128 grid.
A diagonal weighted block choice scheduler policy was chosen that selected block centers from the set all unresolved grid cells weighted by their Euclidean distance
to the upper left corner of the grid.

From the local, nearest neighbor pairwise tile constraints,
the \textit{Forest Micro} tile set has an implied global constraint that the river count from the top row of the realized
grid must match the river count on the bottom row.
This global constraint is not explicitly present or specified but is a by-product of the local constraints.

For large grid sizes, POMS fails to find realizations for the \textit{Forest Micro} tile set when a block
choice policy chooses block centers at random.
Under these conditions, POMS effectively makes a random choice for the number of rivers on the top and bottom row.
One can expect, with enough random sampling, the river count to be identical, but the problem
becomes increasingly less likely as grid size grows.

To guide POMS in finding solutions for the \textit{Forest Micro} tile set, a
block choice policy was chosen that preferentially weights cell positions starting from the upper left and decreases as
cells are considered going down and to the right.
The diagonal weighting has the effect of keeping a growing contiguous region that locally keeps the top and bottom
river counts the same.
Contradictions that do occur tend to be localized and their resolution keeps the river counts the same

Though the \textit{Forest Micro} tile set has an unbounded correlation length, the global constraint is weak enough
to be overcome by this simple weighted diagonal heuristic.

The tile constraint set for 0x72's \textit{Two Bit Micro Metroidvania} (2BMMV) tile set was
generated from the exemplar image shown (Figure \ref{fig:2dexamples}, label \textit{k}) with a 24x24 pixel tile size, a 2x2 super tile window
using a 1x2 tile overlap and taking the upper left hand corner tile as the representative display tile.

As can be seen by Figure \ref{fig:2dexamples} (label \textit{j}), the 2BMMV tile set has unbounded correlation length, but the constraint is weak
enough to be overcome by running POMS with a block size of 48x48 and a block choice scheduler that chooses block centers from
unresolved grid cell positions uniformly.
An example output for the 2BMMV for a 128x128 grid using a 48x48 block size and uniform block choice schedule can be seen in Figure \ref{fig:2dexamples} (label \textit{l}).
The unbounded correlation length comes from the boundary restrictions which might disappear if care were taken
to create wrap around tile constraints.

\subsection{Unbounded Correlation Length but Bounded TACCL}

The 3D \textit{Brutal Plum} tile set was programmatically generated from a set of 20 unique tiles that, after rotations and deduplication,
grows to 90 tiles.
Many of the tiles are aesthetically identical but are logically different to embed desired features, such as requiring certain
logical tiles to be above or below other tiles.
In particular, non empty tiles must have a non empty path to the ground base plane, giving an implied global constraint.
The global constraint is not captured by the Tile Arc Consistent Correlation Length (TACCL) and highlights
the failure of the TACCL heuristic to properly capture an unbounded correlation length embedded in the tile set.

Though the correlation length for the \textit{Brutal Plum} tile constraints is unbounded, POMS still is able to reliably find realizations with
block size 22x22x22, an example output of which is highlighted in Figure \ref{fig:brutal_plum} (label \textit{c}).
For aesthetic reasons, a tile weighting that increases the likelihood of the empty tile, the arch tiles and
stair tiles was chosen.
The reader is referred to the reference implementation \footnote{ \label{poms-url} \url{https://github.com/zzyzek/PunchOutModelSynthesis}. }
for further details.

\section{Conclusion}

Punch Out Model Synthesis (POMS) provides an algorithm for the Constraint Based Tiling Generation (CBTG) problem.
We have shown that POMS can discover realizations from tile constraints that have finite correlation length.
We have also shown that POMS is able to find realizations for some problems
that have weak implied global constraints.
If the tile set and configuration are not conditioned well,
POMS may fail to find a solution or provide biased results.

Tile constraints that have unbounded or long range correlations are more difficult and sometimes intractable.
Constraint Based Tiling Generation problems are, in general, NP-Complete, so there is likely
no comprehensive strategy that leads to efficient methods of solution but the worst case complexity
results sometimes obscure when problems are readily solvable.
For many CBTG problems that are NP-Complete, the general complexity result might not
apply to some generic configuration, allowing some problem ensembles
to be easily solvable.

For many real world CBTG problems, we still lack understanding of the
interplay between how difficult it is to find realizations given tile constraints
and initial configuration.
The Tile Arc Consistent Correlation Length (TACCL) is one heuristic measure that attempts
to quantify how difficult tile constraints are to resolve.
The TACCL has the benefit of being easy to calculate but its interpretation is easily confounded,
so should be considered a coarse measure with limited applicability.

A libre/free reference implementation for Punch Out Model Synthesis (POMS) has been developed and can be downloaded
from its repository \textsuperscript{ \ref{poms-url} }.


\appendix


\section{Appendix: Breakout Model Synthesis (BMS)}

%

\begin{algorithm}[H]
  \caption{Breakout Model Synthesis}
  \label{alg:bms}
  \begin{algorithmic}
    \State \textbf{Input,Output:} Block $B$,
    \State \textbf{Input:} Setup restrictions $S$,
    \State \textbf{Input:} Tile constraints $C$,
    \State \textbf{Input:} Tile Choice Scheduler $TCS$
    \State \textbf{Input:} Soften Choice Scheduler $SCS$

		\State Put $B$ into a fully indeterminate state
		\State Apply setup restrictions $S$ to $B$
    \State Apply tile constraints $C$ until $B$ is arc consistent

		\If{ unable to create initial arc consistent state }
			\State return failure
		\EndIf

		\State Save copy of $B$ to $P$
    \While{ $B$ not fully resolved }
			\State $B' = B$
      \State Choose tile and cell to resolve in $B$ from $TCS$
      \State Propagate resolution until $B$ is arc consistent
      \If{ contradiction encountered }
				\State Revert $B$ back to $B'$
				\State Query $SCS$ for a sub-region, $R$, to soften
				\State Revert region $R$ in $B$ back to $P$
        \State Constraint propagate until $B$ is arc consistent
			\EndIf

      \If{ iteration count has been exceeded }
        \State return failure
      \EndIf
    \EndWhile
    \State return success
  \end{algorithmic}
\end{algorithm}

\end{document}